\documentclass{PoS}
\PoS{PoS(HEP2005)185}
\title{Maximal Neutrino Mixing from Discrete Symmetry in Extra Dimensions}
\ShortTitle{Maximal Neutrino Mixing and $A_4$ Symmetry}
\author{\speaker{Ferruccio Feruglio}\\
        University of Padova, Italy\\
        E-mail: \email{feruglio@pd.infn.it}}
\abstract{
I review the construction of a model for lepton masses based
on the flavour symmetry group $A_4\times$ U(1) reproducing
the so-called tri-bimaximal lepton mixing scheme, in eccelent agreement with
current data. The model predicts a neutrino spectrum 
of normal hierarchy type, not far from degenerate.
A testable relation between neutrino masses is obtained.
I shortly discuss also general requirements for 
models based on spontaneously broken flavour symmetries,
in order to get a maximal atmospheric mixing angle.  

\vskip 1.0 cm 
\centerline{*~~~~~~~*~~~~~~~*}
\vskip 1.0 cm

The present allowed range of leptonic mixing angles
is rather constrained by the available data on  
neutrino oscillations. At the 2$\sigma$ level (95\% C.L.) \cite{data}:
\be
\sin^2\theta_{23}=0.44\times (1^{+0.41}_{-0.22})~~~,~~~~~
\sin^2\theta_{13}=0.9^{+2.3}_{-0.9}\times 10^{-2}~~~,~~~~~
\sin^2\theta_{12}=0.314\times (1^{+0.18}_{-0.15})~~~.
\ee
This range is fully compatible with the so-called Harrison-Perkins-Scott
(HPS) or tri-bimiximal mixing scheme \cite{hps}:
\be
U_{HPS}= \left(\matrix{
\dd\sqrt{\frac{2}{3}}&\dd\frac{1}{\sqrt 3}&0\cr
-\dd\frac{1}{\sqrt 6}&\dd\frac{1}{\sqrt 3}&-\dd\frac{1}{\sqrt 2}\cr
-\dd\frac{1}{\sqrt 6}&\dd\frac{1}{\sqrt 3}&\dd\frac{1}{\sqrt 2}}\right)~~~~~. 
\label{hps}
\ee
Today the errors on $\theta_{~23}$ and $\theta_{13}$ are large and
sizeable deviations from the HPS scheme are still allowed.
{(\em Continue)}}
\FullConference{International Europhysics Conference on High Energy Physics\\
                 July 21st - 27th 2005\\
                 Lisboa, Portugal}
\def\be{\begin{equation}}
\def\ee{\end{equation}}
\def\bc{\begin{center}}
\def\ec{\end{center}}
\def\bea{\begin{eqnarray}}
\def\eea{\end{eqnarray}}
\def\dd{\displaystyle}

\begin{document}
More impressive is the agreement on the $\theta_{~12}$ angle,
the most precisely measured. The HPS scheme predicts $\theta_{~12}\approx 
35.3^0$
while the experimental value is 
$\theta_{~12}=(34.1^{+1.7}_{-1.6})^0$ (1$\sigma$ errors). 
In a not so far future, 
probably less than 10 years from now \cite{schwetz},
also $\theta_{~23}$ and $\theta_{13}$ might be known or constrained
with a comparable precision, thus confirming or excluding the HPS
scheme at the $\lambda^2$ level, $\lambda\approx 0.22$ being
the Cabibbo angle, here regarded as a typical expansion parameter
for flavour mixing. Since the HPS scheme requires a maximal $\theta_{~23}$ 
angle and a vanishing $\theta_{13}$,
two features that are far from generic in model building,
it is interesting to see if it can be justified on the basis of some
dynamical or symmetry principle.

Maximal and vanishing mixing angles are rather special and
we might expect that they arise in the context of a flavour symmetry
in the limit of exact symmetry, that is by neglecting all the symmetry breaking
effects that are typically needed to reproduce detailed features of 
a realistic pattern of fermion masses. 
However this is not the case, at least for realistic
flavour symmetries, where breaking terms are small compared to the leading 
ones \cite{paris04}. Indeed, if the flavour symmetry is broken only by small
effects, the mass matrices for charged leptons and neutrinos 
can be written as:
\be
m_e=m_e^0+...~~~,~~~~~~~~~~m_\nu=m_\nu^0+...
\ee
where dots denote symmetry breaking effects and $m_e^0$
has rank less or equal than one.
Rank greater than one, as for instance when both
the tau and the muon have non-vanishing masses in the symmetry limit,
is clearly an unacceptable starting point, since the difference
between the two non-vanishing masses can only be explained by
large breaking effects.
If the rank of $m_e^0$ vanishes, than all mixing angles in the charged
lepton sector are undetermined in the symmetry limit and
$\theta_{~23}$ is also completely undetermined.
If $m_e^0$ has rank one, then by a unitary transformation we can 
always go to a field basis where
\be
m_e^0=
\left(
\begin{array}{ccc}
0&0&0\cr
0&0&0\cr
0&0&m_\tau^0
\end{array}
\right)~~~.
\ee
Denoting by $U_\nu$ and $U_e$ the unitary matrices that diagonalize $m_\nu^0$ and 
$m_e^{0\dagger} m_e^0$, we have
\be
U_e=R_{12}(\theta^e_{12})~~~
\ee
where the angle $\theta^e_{12}$ is completely undetermined
($R_{ij}$ is the orthogonal matrix representing a rotation in the
$ij$ sector). Moreover, by neglecting phases and adopting the standard 
parametrization
$U_\nu=R_{23}(\theta^\nu_{23}) R_{13}(\theta^\nu_{13}) 
R_{12}(\theta^\nu_{12})$, we find that the angle $\theta_{~23}$
of the physical mixing matrix $U_{PMNS}=U_e^\dagger U_\nu$
is given by:
\be
\tan\theta_{23}=\cos\theta^e_{12} \tan\theta^\nu_{23}
+
\sin\theta^e_{12}\frac{\tan\theta^\nu_{13}}{\cos\theta^\nu_{23}}
~~~.
\label{tan23}
\ee
Therefore, in general, the atmospheric mixing angle is always
undetermined at the leading order (this conclusion is unchanged if phases
are accounted for). When small symmetry breaking
terms are added to $m_e^0$ and $m_\nu^0$, it is possible to
obtain $\theta_{~23}=\pi/4$, provided these breaking
terms have suitable orientations in the flavour space.

If the breaking terms originate from a spontaneous symmetry breaking,
there are four requirements to satisfy in order to obtain
the HPS scheme. 1) Two independent scalar sectors are needed.
One of them communicates the breaking to charged fermions
and the other one feeds the breaking to neutrinos.
In such a framework a maximal atmospheric mixing angle is always
the result of a special vacuum alignment between these two sectors.
2) This alignment should be natural. It should correspond to
a local minimum of the potential energy of the theory,
in a finite region of the parameter space, {\em i.e.} without
enforcing any ad-hoc relation among parameters.
3) 
The alignment should not be spoiled by large sub-leading terms.
In general the mixing angles are power series in the symmetry breaking
order parameters. Calling $\langle \varphi \rangle$ the 
generic such parameter, even by enforcing HPS at the leading order, we have:
\be
\theta_{13}=0+a_1\dd\frac{\langle \varphi \rangle}{\Lambda}
+a_2 \dd\frac{\langle \varphi \rangle^2}{\Lambda^2}+...~~~,~~~~~~~~~~
\theta_{23}=\dd\frac{\pi}{4}+b_1\dd\frac{\langle \varphi \rangle}{\Lambda}
+b_2 \dd\frac{\langle \varphi \rangle^2}{\Lambda^2}+...
\ee
the higher-order corrections coming from higher-dimensional operators
compatible with the flavour symmetry.
It is not sufficient that the alignment produces the desired 
first term in the expansion. We should also be able to keep
under control the remaining contributions. This can be done
either by adopting, if possible, a small breaking parameter 
(for instance $\langle \varphi \rangle/\Lambda<\lambda$), or
by building the model in such a way that the first corrections
$a_1$ and $b_1$ vanish. 
4) Finally, the alignment should be compatible with the mass hierarchies.
In particular $m_e/m_\tau$ and $m_\mu/m_\tau$ should vanish
when $\langle \varphi \rangle/\Lambda$ is set to zero.

A close relation between the HPS scheme and the discrete symmetry group
$A_4$ has been known for some time \cite{a4}
and, indeed, all the requirements listed above can be
satisfied in a model for lepton masses 
based on the flavour symmetry $A_4\times$ U(1) \cite{af},
where the U(1) factor controls the charged lepton mass hierarchies.
The group $A_4$ is made of the twelve three-dimensional rotations 
leaving invariant a tetrahedron and possesses four representations: 
three singlets $1$, $1~'$, $1''$ and a triplet 3. 
The assignment of the relevant fields to representations of $A_4\times $ U(1)
is given in the table,
\begin{center}
\begin{tabular}{|c|c||c|c|c||c|c|c|c|c|}
\hline
{\tt Field}& l & $e^c$ & $\mu^c$ & $\tau^c$ & $h_{u,d}$ & 
$\varphi$ & $\varphi'$ & $\xi$ & $\theta$\\
\hline
$A_4$ & $3$ & $1$ & $1'$ & $1''$ & $1$ & 
$3$ & $3$ & $1$ & $1$\\
\hline
$U(1)$ & $0$ & $4$ & $2$ & $0$ & $0$ & 
$0$ & $0$ & $0$ & $-1$\\
\hline
\end{tabular}
\end{center}
where the symmetry breaking sector is described in the last five
columns. The VEV of $\theta$ breaks the U(1) symmetry and provides
the correct hierarchy to charged leptons. The alignment needed
to reproduce the HPS pattern is 
\be
\langle \varphi' \rangle=(v',0,0)~~~,~~~~~~~~ \langle \varphi \rangle=(v,v,v)~~~,
~~~~~~~\langle \xi \rangle=u~~~, 
\label{align}
\ee
where, in the absence of particular relations among the parameters
of the model, $v\approx v~'\approx u\approx \langle\theta\rangle$
is expected. A simple, not necessarily unique, set up that gives rise
to (\ref{align}) is depicted in the figure.
\begin{figure}[h!]
\begin{center}
\includegraphics[width=0.4\textwidth]{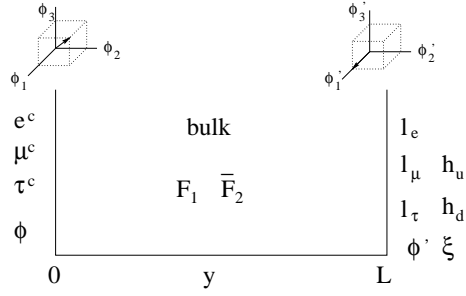}
\caption
{Fifth dimension and localization of scalar and fermion fields.
The symmetry breaking sector includes the $A_4$ triplets $\varphi$
and $\varphi'$, localized at the opposite ends of the interval.
Their VEVs are dynamically aligned along the directions shown
at the top of the figure.}
\end{center}
\end{figure}
The fields $\varphi$ and $(\varphi',\xi)$, giving masses respectively
to charged leptons and to neutrinos, live at the opposite ends
of an extra spatial dimension and it can be shown that this naturally leads 
to the desired alignment. Left-handed
leptons live on the brane at $y=L$ and neutrinos acquire masses
directly from the operators:
\be
\frac{1}{\Lambda^2}\xi (ll)h_u h_u~~~,~~~~~~~~~~~~~~~~~~
\frac{1}{\Lambda^2} (\varphi' ll)h_u h_u~~~.
\ee
Right-handed leptons live in $y=0$ and $e$, $\mu$ and $\tau$ 
get their masses indirectly, by the exchange of an heavy bulk 
fermion $F$ of mass $M$ that interacts on the two branes through
the operators:
\be
\dd\frac{(f^c \varphi F)}{\sqrt{\Lambda}}\delta(y)~~~,~~~~~~~~~~
\dd\frac{(F^c l)h_d}{\sqrt{\Lambda}}\delta(y-L)~~~.
\ee
At energies much smaller than $M$, an effective Yukawa of the kind
$(f^{~c}\varphi l) h_d e^{-M L}/\Lambda$ is generated.
At leading order, by neglecting possible higher dimensional operators,
the HPS scheme and the correct hierarchies of charged fermion masses
are obtained. A detailed analysis, including possible non-leading
effects arising from higher dimensionality operators, reveals 
that the first corrections to the HPS mixing pattern 
only arise at the second order in the expansion parameter $VEV/\Lambda$.
In this model, in order to accommodate the hierarchy of charged fermion masses, 
$VEV/\Lambda\approx \lambda\approx 0.22$ and thus the expected deviations
from the HPS scheme are tiny. 
The neutrino spectrum is of normal hierarchy type and, 
in a large portion of the parameter space, we find
$|m_{~3}|\approx 0.053~~{\rm eV}$, $|m_1|\approx |m_2|\approx 0.017~~{\rm eV}$.
An accidental quasi-degeneracy is not excluded. 
We also have $|m_{ee}|\approx 0.005$ eV, at the upper edge of
the range allowed for normal hierarchy, but unfortunately too small
to be detected in a near future. In the whole parameter space, barring
small corrections, the following testable relation holds:
\be
|m_3|^2=|m_{ee}|^2+\frac{10}{9}\Delta m^2_{atm}\left(1-
\frac{\Delta m^2_{sol}}{2\Delta m^2_{atm}}\right)~~~.
\ee
This model provides an existence proof of a class of `special' models
where, at variance with most of the existing models \cite{review}, $\theta_{~23}$ is maximal
and $\theta_{13}$ is vanishing within tiny corrections which are probably below the 
sensitivity achievable in the near future.

\end{document}